\documentclass[twocolumn]{aastex631}
\usepackage{xcolor}

\shorttitle{Testing the $\tau$-$M_{BH}$ linear relationship}
\shortauthors{Chen et al.}
\graphicspath{{./}{figures/}}

\begin{document}
\title{Testing the linear relationship between black hole mass and variability timescale in low-luminosity AGN at submillimeter wavelengths}
%

\author[0000-0002-1633-8271]{Bo-Yan Chen}
\affiliation{Academia Sinica Institute of Astronomy and Astrophysics, No.1, Sec. 4, Roosevelt Rd, Taipei 10617, Taiwan, R.O.C.}

\author[0000-0003-4056-9982]{Geoffrey C.\ Bower}
\affiliation{Academia Sinica Institute of Astronomy and Astrophysics, 645 N. A'ohoku Place, Hilo, HI 96720, USA}

\author[0000-0003-3903-0373]{Jason Dexter}
\affiliation{JILA and Department of Astrophysical and Planetary Sciences, University of Colorado, Boulder, CO 80309}

\author[0000-0001-9564-0876]{Sera Markoff}
\affiliation{Anton Pannekoek Institute for Astronomy, University of Amsterdam, Science Park 904, 1098 XH, Amsterdam, The Netherlands}

\author{Anthony Ridenour}
\affiliation{W. M. Keck Observatory, 65-1120 Mamalahoa Hwy., Kamuela, HI 96743, USA}

\author[0000-0003-0685-3621]{Mark A. Gurwell}
\affiliation{Center for Astrophysics | Harvard $\&$ Smithsonian, 60 Garden Street, Cambridge MA 02138 USA}

\author[0000-0002-1407-7944]{Ramprasad Rao}
\affiliation{Center for Astrophysics | Harvard $\&$ Smithsonian, 60 Garden Street, Cambridge MA 02138 USA}

\author[0000-0001-8345-7097]{Sofia H. J. Wallström}
\affiliation{Institute of Astronomy, KU Leuven, Celestijnenlaan 200D, 3001 Leuven, Belgium}

\begin{abstract}
The variability of submillimeter emission provides a useful tool to probe the accretion physics in low-luminosity active galactic nuclei. 
We accumulate four years of observations using Submillimeter Array for Centaurus A, NGC 4374, NGC 4278, and NGC 5077 and one year of observations for NGC 4552 and NGC 4579. 
All sources are variable. 
We measure the characteristic timescale at which the variability is saturated by modeling these sources' light curve as a damped random walk. 
We detect a timescale for all the sources except NGC 4552. 
The detected timescales are comparable to the orbital timescale at the event horizon scale for most sources. 
Combined with previous studies, we show a correlation between the the timescale and the black hole mass over three orders of magnitude. 
This discovery suggests the sub-mm emission is optically thin with the emission originating from the event horizon. 
The mass scaling relationship further suggests that a group of radio sources with a broadband spectrum that peaks at submillimeter wavelengths have similar inner accretion physics. 
Sources that follow this relationship may be good targets for high-resolution imaging with the Event Horizon Telescope.

\end{abstract}
\keywords{Active galactic nuclei: Low-luminosity AGN - Active galactic nuclei: Radio active galactic nuclei - Active galactic nuclei: Radio core - Black holes: Supermassive black holes - Astronomical methods: Time domain astronomy: Time series analysis}

\section{Introduction} \label{sec:intro}
    
    In recent years, people have shown an increased interest in understanding the similarity of black hole accretion systems. 
    Studies have drawn analogs between systems with mass across several orders of magnitude, from stellar mass black hole in X-ray binaries (BHBs) to supermassive black hole (SMBH) in active galactic nuclei (AGN).
    Long-term variability studies have revealed that the timing feature of the accretion behavior is related to physical parameters like black hole mass, luminosity, Eddington ratio, and redshift (e.g, \citet{2006Natur.444..730M}, \citet{2009ApJ...698..895K}, and \citet{2010ApJ...721.1014M}).
    More specifically, these studies have investigated the relationship between black hole mass and characteristic timescale of variability.
    One can expects a linear relationship between black hole mass and characteristic timescale if the emission origin region in all systems is the same size in the Schwarzschild radius unit ($R_s = 2 GM/c^2$).
    The characteristic timescale, corresponding to the transition in the power spectral density (PSD) of the variability varying from power law in the high-frequency regime to white noise in the low-frequency end, have been found be consistent with the physical timescales, e.g., the orbital timescale or the thermal timescales; therefore, the characteristic timescale corresponds to the emission zone size and is further related to the black hole mass.
    This concept was clearly shown in \citet{2021Sci...373..789B} who suggested that the physical origin of the mass-dependent optical timescale is associated with the thermal timescale.
    The corresponding emission radius of the optical timescale deriving from the thin disk theory (\citet{1973A&A....24..337S} is consistent with the microlensing measurement (see e.g., \citet{2010ApJ...712.1129M} and \citet{2018ApJ...869..106M}).
    Further, a comparison with the substantially shorter mass-dependent x-ray timescales showed the potential of that variability is able to probe different scale of accretion disk by selecting different wavelength.
    
    In addition, studies of the broad band spectrum have suggested a strong correlation between radio, x-ray, and black hole mass (so call the "fundamental plane of black hole activity" or FP, see \citet{2003MNRAS.345.1057M} and \citet{2004AA...414..895F}). 
    This relationship confirms the prediction of earlier theoretical framework, in which the synchrotron spectrum feature from a relativistic jet is expected to scale with the accretion power and the black hole mass (\citet{1979ApJ...232...34B} and \citet{1995AA...293..665F}).
    Therefore, the Sub-Eddington systems, whose spectral energy distribution (SED) is jet-dominated, show a tight relationship between the optically thick radio emission and the optically thin optical or x-ray emission when the mass and accretion power is scaling along with black hole systems.
    Collectively, the black hole systems, regardless of disk-dominated (corresponding to the standard thin disk accretion mode) or jet-dominated (corresponding to the optically thin and radiatively inefficient accretion mode; see \citet{2014ARAA..52..529Y}), seems to have a mass-scale-free characteristic.

    In previous study, \citet{2015ApJ...811L...6B} investigated the relationship between timescale and black hole mass at submillimeter (hereafter sub-mm) wavelengths.
    Two sources’ timescale, M81 and M87, were detected, in contrast to only a lower limit constraint on most of the source in the SMA calibrator database.
    Combining with previous Sgr A*’s measurement (\citet{2014MNRAS.442.2797D}), they revealed a linear timescale-black hole mass correlation among three low-luminosity AGNs (LLAGNs; \citet{2008ARAA..46..475H}).
    Three measured timescales, which all corresponding to the physical timescale at event-horizon scale radius in their system, as well as the linear relationship indicate that the variability origin from the black hole vicinity where the sub-mm emission is optically thin.
    This interpretation can be supported by these galaxies’ multi-wavelength study, which they all peak in the sub-mm band while showing similar spectrum shape in radio to sub-mm wavelength (\citet{2008ApJ...681..905M}); further, with their very long baseline interferometry (VLBI) experiment result (\citet{2019ApJ...875L...1E}, \citet{2022ApJ...930L..12E}), these evidence suggest three LLAGNs’ inner part accretion physics may be similar.
    Other LLAGNs, having similar radio properties, are also expected to follow this mass-timescale relationship in the sub-mm band.
    
    Understanding sub-mm variable behavior is crucial for studying the astrophysics of these LLAGN systems.
    In their comprehensive investigation of Sgr A*’s sub-mm light curve using high SNR, high cadence ALMA phased array data as the byproduct of the 2017 EHT campaign, \citet{2022ApJ...930L..19W} showed that the variability of 230 GHz emission in a short timescale from one minute to several hours can be characterized by a red noise process. 
    Furthermore, the PSD slope on minutes to hours timescale is steeper than the generally used damped random walk model (DRW model; see \citet{2009ApJ...698..895K}).
    Moreover, a comparison between observational variability and numerical simulated variability, done by \citet{2022ApJ...930L..16E}, showed that the simulation, however, seems to produce exceed variability. 
    In other words, the measured variability can provide a strong constraint on the simulated models; nevertheless, the discrepancy could also be attributed to some unknown physical processes which may suppress the emission variation.

    Motivated by sub-mm light curve studies mention above, in this paper, we attempt to extend the variability study to nearby bright LLAGNs, and we intend to explore the black hole mass-characteristic timescale correlation in sub-mm wavelengths.
    Additionally, this study provides potential candidates for sub-mm high angular resolution imaging.
    The paper is structured as follows.
    We present the SMA observation and light curves, including archive data in section \ref{sec:lc}. 
    We then apply two timescale analysis methods in section \ref{sec:tau_anal} and discuss the mass scaling relationship result in section \ref{sec:discu}. 
    Finally, in section \ref{sec:conclu}, we give our research conclusions.
    
\section{Observational Data} \label{sec:lc}
    \subsection{SMA observation}
        We designed a monitoring observation toward six nearby LLAGNs on a small patch of sky (RA $\sim$ 1 hr) in a weekly to daily observation frequency.
        Six LLAGNs, Centaurus A (hereafter Cen A), NGC 4374, NGC 4552, NGC 4579, NGC 4278, and NGC 5077, were selected.
        These sources are predicted to have variability timescales ranging from a few days to ten days, due to their black hole mass being situated between Sgr A* and M87. 
        This is expected to provide clarity on the sub-millimeter mass scaling relationship.
        In addition, these sources are bright in sub-mm wavelength ($\geq$ 25 mJy) and show a similar radio properties to Sgr A*, M81, and M87, that is, a flat to inverted spectrum from centimeter to sub-mm wavelength (\cite{2011AJ....142..167D}).
        
        Six targets were monitored using SMA from June 2015 to January 2019.
        We adopted the "piggy-backed" observation on the full-track schedule to collect light curve; therefore, observational frequency varied around 230 GHz.
        Note that the angular resolution varies from $\sim$1 to $\sim$10 arcsecond, which is related to the array configuration schedule.
        The observations were done under good weather conditions with precipitation water vapor (PWV) $<$ 4mm.
        
        We performed initial flagging and calibration, including bandpass, flux, and gain calibration in IDL software MIR.
        We then performed imaging and flux measurements in the Miriad package.
        We carried out flux calibration using planet moon, mainly Callisto, Titan, or Ganymede.
        Bright sources such as 3C 279, 3C 454.3, or 3C 84 were used for the bandpass calibration.
        The choice of flux and bandpass calibrator depends on the available source in the sky.
        We performed gain calibration with calibrator 1315-336 for Cen A, 1159+292 for NGC4278, 1305-105 for NGC 5077, and 3C 274 for NGC 4374, NGC 4552, and NGC 4579.
        After the calibration, each receiver and sideband were imaged separately.
        Source detection was confirmed by both visual and a three SNR threshold applying on individual image.
        Once the detection was confirmed, we applied a point-source model fit on the visibility.
        The measurement of flux and uncertainty from the visibility was then adopted into the light curve.
        Fitting flux directly on the visibility gets rid of the extra uncertainty that may be introduced by the imaging process.
        The light curve data points will be filtered out when a noticeable mismatch in the calibrator's flux, which cannot be explained by the calibrator's variability, is detected between the measurement and the archive.
        All the targets in the individual map showed a point-source-like structure, and no extended emission has been detected.
        
    \subsection{The archive data}
        In addition to the monitoring observation, we compiled data from the SMA calibrator database \footnote{\url{http://sma1.sma.hawaii.edu/callist/callist.html}} and ALMA archive for M87, Cen A, NGC 4552, and NGC 4579.
        In the cases of M87 and Cen A, the SMA calibrator database includes observations from April 2003 and July 2005 to the present, spanning about 18 and 15 years, respectively.
        We adopted NGC 4552 and NGC 4579’s ALMA observation from the JVO image archive \footnote{\url{https://jvo.nao.ac.jp/portal/alma.do}}. 
        We measured the flux density from the FITS images in the JVO image archive by modeling the source as a point source, and we estimated the uncertainty by taking the rms value from the off-source region.
        Furthermore, we included a measurement of NGC 4579 with a flux of $11 \pm 2$ mJy observed by PdBI (\citet{2007A&A...464..553K}) around 2002.
        These archive data significantly expand the light curve duration and provide useful information for understanding the long-term flux level of our targets.
        
    \subsection{Light Curves}
\begin{figure*}[ht!]
            \centering
            \includegraphics[width=\textwidth]{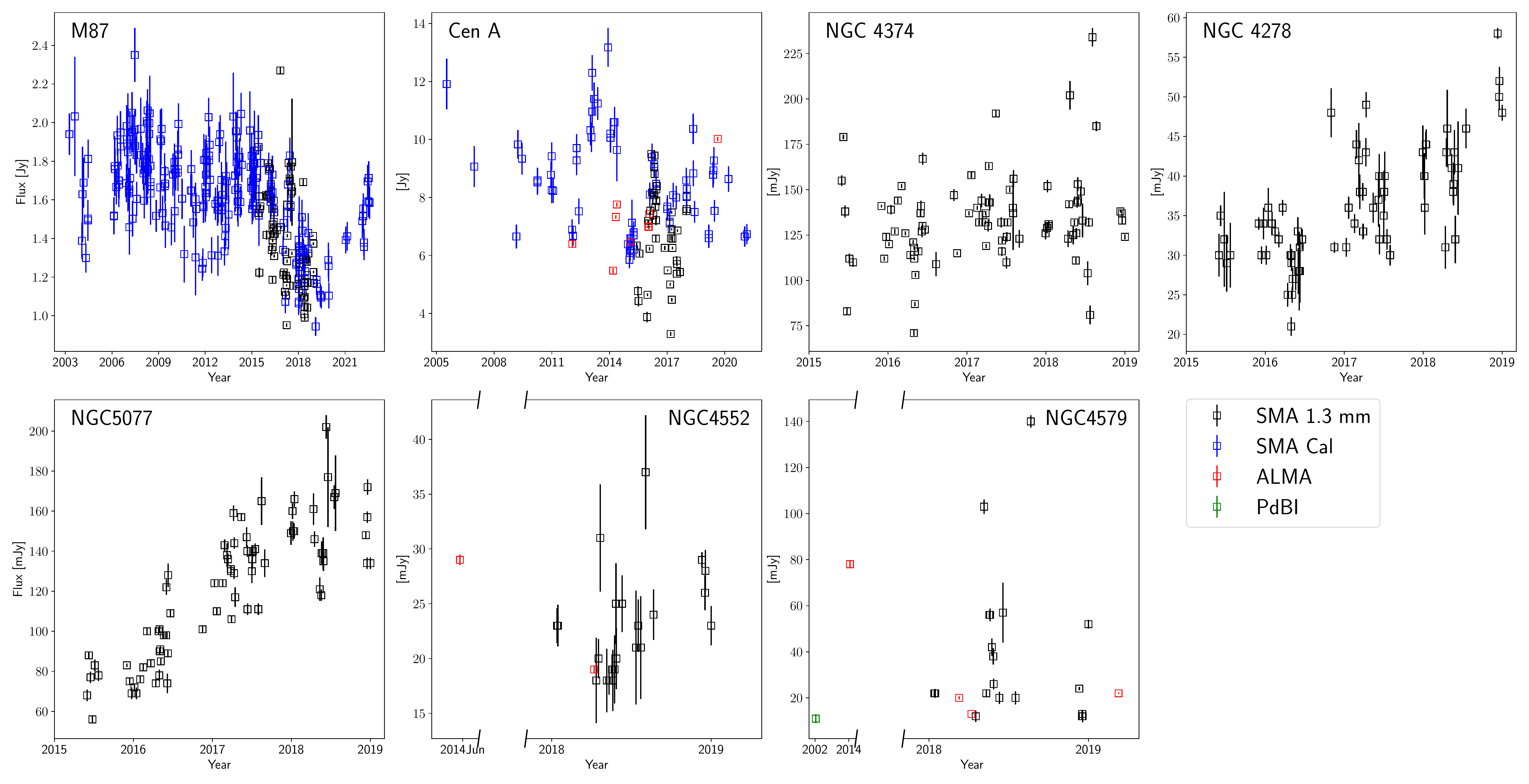}
            \caption{Light curves of M87, Cen A, NGC 4374, NGC 4278, NGC 5077, NGC 4552, and NGC 4579. The data point in black labels our SMA observation, while blue and red represents the SMA calibrator list and ALMA archive data. The data point in green shows the PdBI measurement for NGC 4579. We remove the large gap in NGC 4552 and NGC 4579.}
            \label{fig:lc}
        \end{figure*}

        Light curves of M87, Cen A, NGC 4374, NGC 4278, NGC 5077, NGC 4552, and NGC 4579 are shown in Figure \ref{fig:lc}.
        We use symbols with different colors to denote data from either our observation or archive.
        The full light curve's information are shown in Table \ref{table:lcinfo}.
        We present our unpublished SMA data in Table \ref{table:sma}.
        The full electronic version can be downloaded on the journal website.

        The consist of each light curve is shown in Table \ref{table:lcinfo}.
        In addition, we have included M87 (3C 274) in this study.
        Variability properties of M87 was investigated in \citet{2015ApJ...811L...6B} by using a 11-years light curve with 147 data points.
        In this paper, we have extended the time baseline to 18 years with 314 data points.
        We further included M87’s flux measurement from our observation, which provides a finer cadence for probing M87's variability.
        
        Cen A and NGC 4374 show long-term stability, which is consistent with our understanding of LLAGN variability property (\citet{2014MNRAS.442.2797D} and \citet{2015ApJ...811L...6B}).
        NGC 4552 and NGC 4579 also show stability during their whole light curve; however, note that we only have a small number of data across a short period.
        This suggests that the result in the following analysis section should be tread cautiously.
        On the other hand, flux levels gradually increase for NGC 4278 and NGC 5077, suggesting that these two sources may behave more like a regular AGN.
        For M87, we observed a long declining trend starting from 2015, but have reversed itself and reached the mean again recently, which was not included in the previous study, making it worth re-analyzing.

        \begin{table*} \footnotesize
        \caption{Information of full light curve.}
        \begin{tabular}{c c c c c c}
        \hline
        Source & Start Date & End Date & Consist & Mean Flux & Var\tablenotemark{a}\\
         & [MJD] & [MJD] & [$\#$] & [Jy] & [$\%$]\\
        \hline
        M87      & 52736.0 & 59789.0 & SMA: 83, SMACAL: 231         & 1.571 $\pm$ 0.261 & 15\\
        Cen A    & 53564.0 & 59267.0 & SMA: 44, SMACAL: 61, ALMA: 9 & 7.743 $\pm$ 1.766 & 22\\
        NGC 4278 & 57174.0 & 58484.0 & SMA: 69                      & 0.036 $\pm$ 0.007 & 19\\
        NGC 4374 & 57174.0 & 58484.0 & SMA: 83                      & 0.133 $\pm$ 0.024 & 18\\
        NGC 4552 & 56803.0 & 58484.0 & SMA: 22, ALMA: 2             & 0.023 $\pm$ 0.005 & 16\\
        NGC 4579 & 52379.0 & 58553.0 & SMA: 19, ALMA: 5, PdBI: 1    & 0.037 $\pm$ 0.031 & 84\\
        NGC 5077 & 57174.0 & 58484.0 & SMA: 75                      & 0.120 $\pm$ 0.033 & 27\\ 
        \hline
        \end{tabular}
        \tablenotetext{a}{The mean fraction of variation is defined as Var$=\sqrt{\sigma^2 - \delta^2}/\langle F \rangle$ where $\sigma^2$ is the variance of the light curve, $\delta^2 = 1/N\sum^N_i\delta_i^2$ is the mean square of flux uncertainties, and $\langle F \rangle$ is the mean of flux.}
        \label{table:lcinfo}
        \end{table*}

        \begin{table*}
        \centering
        \caption{Table of unpublished SMA data. Ellipsis in the table represents the whole data that we do not display here. The full version can be downloaded on the e-journal website.}
        \begin{tabular}{c c c c c c}
        \hline
        Source & Epoch & Freq & Flux Density & Beam Size & PA\\
         & [yyyymmdd] & [GHz] & [Jy] & [arcsec$^2$] & [degree]\\
        \hline
        M87 & 20150601 & 224.5 & 1.532 $\pm$ 0.018 & 3.92$\times$2.09 & -42.2\\
        M87 & 20150609 & 224.5 & 1.874 $\pm$ 0.009 & 7.38$\times$4.24 & 72.8\\
        M87 & 20150617 & 224.5 & 1.618 $\pm$ 0.018 & 9.21$\times$3.78 & 60.3\\
        M87 & 20150626 & 225.0 & 1.223 $\pm$ 0.026 & 3.83$\times$2.29 & 21.3\\
        M87 & 20150707 & 224.5 & 1.551 $\pm$ 0.017 & 4.83$\times$2.27 & 20.9\\
         & ...\\
        \hline
        CenA & 20150601 & 224.5 & 6.343 $\pm$ 0.101 & 9.13$\times$2.27  & -29.7\\
        CenA & 20150626 & 225.0 & 4.770 $\pm$ 0.172 & 8.33$\times$2.60  & -10.9\\
        CenA & 20150707 & 224.5 & 4.428 $\pm$ 0.192 & 10.65$\times$2.51 & 6.7\\
        CenA & 20150724 & 211.0 & 6.066 $\pm$ 0.101 & 13.40$\times$2.52 & 12.4\\
        CenA & 20151215 & 221.0 & 3.875 $\pm$ 0.184 & 8.47$\times$2.58  & -23.9\\
         & ...\\
        \hline
        NGC4374 & 20150601 & 224.5 & 0.155 $\pm$ 0.003 & 3.85$\times$2.05 & -45.4\\
        NGC4374 & 20150609 & 224.5 & 0.179 $\pm$ 0.001 & 7.37$\times$4.24 & 73.8\\
        NGC4374 & 20150617 & 224.5 & 0.138 $\pm$ 0.003 & 9.16$\times$3.77 & 59.9\\
        NGC4374 & 20150626 & 225.0 & 0.083 $\pm$ 0.002 & 3.82$\times$2.28 & 20.3\\
        NGC4374 & 20150707 & 224.5 & 0.112 $\pm$ 0.003 & 4.84$\times$2.28 & 20.6\\
         & ...\\
        \hline
        NGC4278 & 20150601 & 224.5 & 0.030 $\pm$ 0.003 & 3.60$\times$2.08 & -45.1\\
        NGC4278 & 20150609 & 224.5 & 0.035 $\pm$ 0.001 & 7.76$\times$3.94 & 70.8\\
        NGC4278 & 20150626 & 231.0 & 0.032 $\pm$ 0.006 & 3.76$\times$2.39 & 14.6\\
        NGC4278 & 20150707 & 230.0 & 0.029 $\pm$ 0.004 & 4.37$\times$2.25 & 12.2\\
        NGC4278 & 20150724 & 205.0 & 0.030 $\pm$ 0.004 & 4.88$\times$3.05 & 10.2\\
         & ...\\
        \hline
        NGC5077 & 20150601 & 224.5 & 0.068 $\pm$ 0.003 & 5.53$\times$2.09 & -44.3\\
        NGC5077 & 20150609 & 224.5 & 0.088 $\pm$ 0.001 & 7.92$\times$5.44 & -87.3\\
        NGC5077 & 20150617 & 224.5 & 0.077 $\pm$ 0.002 & 9.84$\times$4.73 & 55.1\\
        NGC5077 & 20150626 & 225.0 & 0.056 $\pm$ 0.002 & 4.33$\times$2.45 & 10.7\\
        NGC5077 & 20150707 & 224.5 & 0.083 $\pm$ 0.003 & 5.69$\times$2.29 & 18.0\\
         & ...\\
        \hline
        NGC4552 & 20180113 & 221.0 & 0.023 $\pm$ 0.002 & 2.56$\times$2.41 & 30.5\\
        NGC4552 & 20180115 & 216.5 & 0.023 $\pm$ 0.002 & 4.02$\times$2.74 & -32.3\\
        NGC4552 & 20180413 & 202.0 & 0.018 $\pm$ 0.004 & 1.32$\times$0.89 & 42.2\\
        NGC4552 & 20180418 & 218.5 & 0.020 $\pm$ 0.002 & 1.04$\times$0.86 & -82.4\\
        NGC4552 & 20180422 & 220.5 & 0.031 $\pm$ 0.005 & 1.70$\times$0.82 & 75.8\\
         & ...\\
        \hline
        NGC4579 & 20180113 & 221.0 & 0.022 $\pm$ 0.002 & 2.57$\times$2.50 & 57.5\\
        NGC4579 & 20180115 & 216.5 & 0.022 $\pm$ 0.002 & 4.14$\times$2.82 & -35.4\\
        NGC4579 & 20180418 & 227.0 & 0.012 $\pm$ 0.002 & 0.98$\times$0.80 & -82.6\\
        NGC4579 & 20180507 & 217.0 & 0.108 $\pm$ 0.007 & 1.46$\times$0.93 & -60.8\\
        NGC4579 & 20180512 & 225.0 & 0.022 $\pm$ 0.001 & 1.56$\times$0.88 & -86.1\\
         & ...\\
        \hline
        \end{tabular}
        \label{table:sma}
        \end{table*}

\section{Timescale analysis} \label{sec:tau_anal}
    
    \subsection{Structure Function} \label{subsec:sf}
\begin{figure*}[h!]
            \centering
            \includegraphics[width=\textwidth]{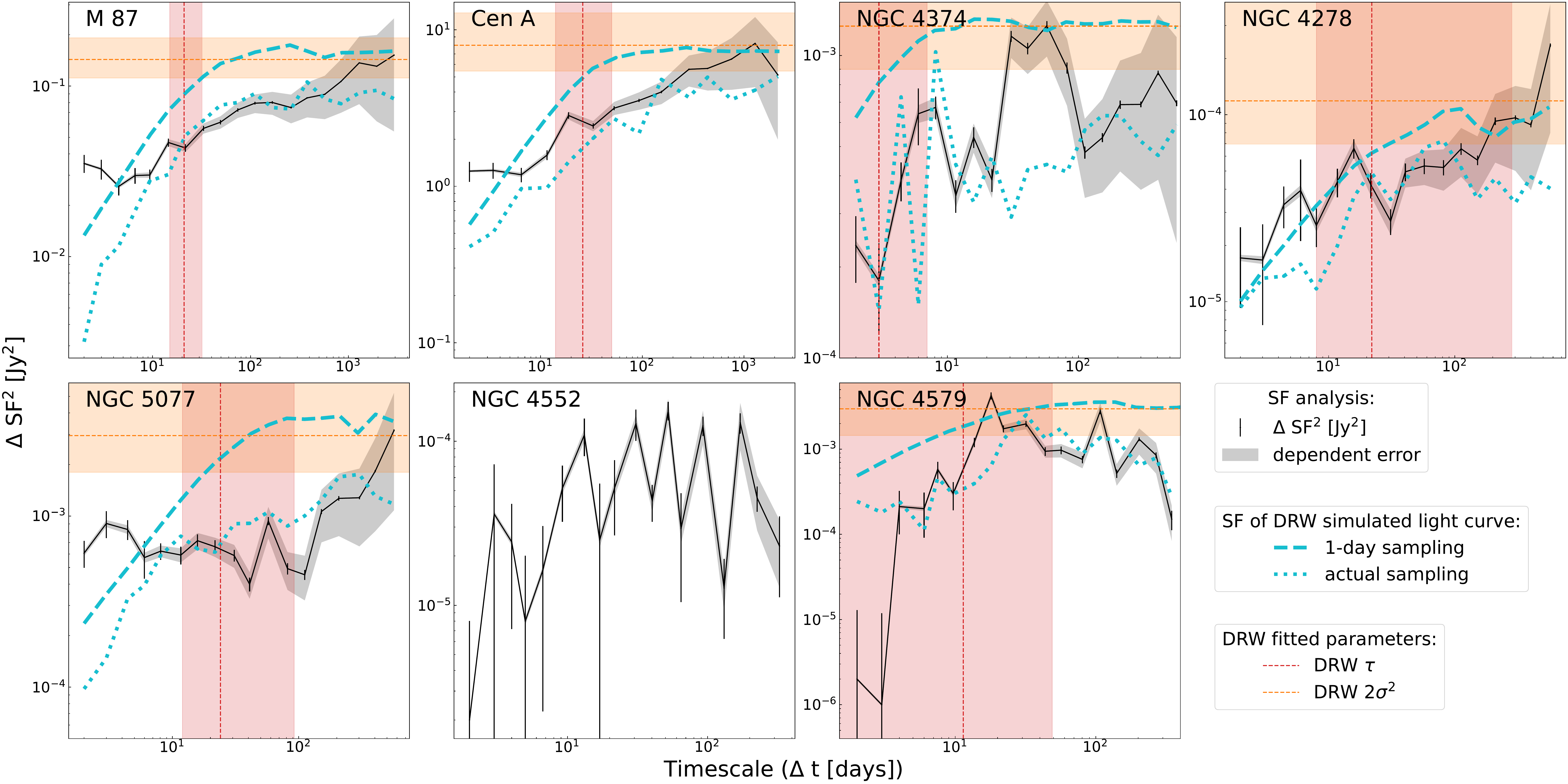}
            \caption{The structure function analysis result. The error bar shows one sigma uncertainty inferred from the measurement noise variance and the sample numbers. The grey shaded area represents the error caused by the finite number of observations contributing to each bin's measurement. We show the simulated DRW light curve with parameter inferred from section \ref{subsec:drw}'s fitting. The cyan dashed line shows the regular sampling of the light curve, while the cyan dotted line shows the actual sampling. The red vertical dashed line indicates the inferred timescale, while the orange horizontal dashed line represents twice the variance, which indicates the expected level of the long-term flux variability for a stationary signal. The comparison of data and simulation suggests that in our cases, SF may be inadequate for accurately identifying the variability timescale.
            }
            \label{fig:sf}
        \end{figure*}

        We have applied a first-order structure function analysis to the light curves shown in the previous section.
        The structure function SF($\Delta t_i$) of a light curve $F_j$ with N numbers of data points in $\Delta t_i$ time bin is defined as
        \begin{equation}
        SF(\Delta t_i)^2 = \frac{1}{N(\Delta t_i)} \sum_j \Delta F_j^2(\Delta t_i).
        \label{eq:sf_d}
        \end{equation}
        The SF results are shown in Figure \ref{fig:sf}.
        Error bars, deriving from calculations in \citet{1985ApJ...296...46S}'s Appendix, show one $\sigma$ of measurement noise variance plus the combined statistic error in each bin.
        However, the SF method have been shown to suffer from the dependent sampling issue, that is, each of the SF point is correlated to others, and the numbers of independent sampling are finite.  
        For example, a 1000 days observation duration can only have five independent measurements for a process that operates on a timescale of 200 days.
        This sampling issue will be more significant at the long timescale.
        In this study, we estimated this uncertainty by considering the numbers of independent measurement ($N_{ind}$) in each bin.
        SF point in each bin then exists a $1/\sqrt{N_{ind}}$ fraction of uncertainty, which is displayed as the gray shade region in Figure \ref{fig:sf}.
        
        SF provides observation on various between different time lag; therefore it can reflect the slope properties of PSD on short timescale as well as detect the variability saturation on long timescale.
        A typical SF is composed by three parts: a flat component in the shortest timescale representing the noise level of the measurement, an increasing trend with a constant slope on a short timescale, and a slope changing to a flat plateau which reflect the saturation of the variability on a long timescale (see e.g. Figure 8 in \citet{2022ApJ...930L..19W}).
        In our SF results, M87, Cen A, and NGC 4374 show an apparent changing of slope on the timescale of ten to hundred days. 
        On the other hand, a suspected steep to flat characteristic is detected in NGC 4278’s SF, while no detection is found for NGC 5077. 
        Lacking of a flat plateau for NGC 4278 and NGC 5077 suggests that there is no stable flux level in their long term variability, as we can observe in their light curve.
        With only a few data points ($\sim$ 20), the SF of NGC 4552 and NGC 4579 are ambiguous.
        Therefore, we believe the saturated timescale of a few to ten days in these cases could be a spurious detection.\\

        SF has been widely applied in light curve variability studies (e.g., \citet{2001ApJ...555..775C}, \citet{2003MNRAS.342.1222C}, \citet{2014MNRAS.442.2797D}, and \citet{2016ApJ...826..118K}) to avoid the ground-based observational bias.
        It provides a useful insight into the qualitative comparison between sources when the actual PSD is relatively difficult to be derived.
        However, \citet{2010ApJS..187..135E} pointed out that the limitation of SF related to three points. 
        First, properly handling the estimation of the timescale and its uncertainty could be challenging because SF data points are non-independent.
        Second, any specific total sampling duration may lead to a spurious saturated timescale.
        Most important, SF can still be affected by the observational gap in the light curve.
        
        We further compare the Gaussian Process (GP) model fitting result (the DRW in the next section) with the SF measurement.
        We present the SF of a simulated DRW light curve with parameters inferred from the DRW fitting process.
        In figure \ref{fig:sf}, the DRW light curve is labeled in cyan.
        The dashed line represents the simulated light curve with a regular sampling cadence of one days, while the dotted line shows the actual sampling of the SMA observation, and both simulated light curves present the actual duration of the real data.
        Additionally, we exhibit the DRW inferred parameters such as the timescale (vertical red dashed line) and the twice of long-term variance (vertical orange dashed line).
        The analysis suggest that for M87 and Cen A, the SF (black line) is consistent with the DRW model (cyan dashed line), and the DRW inferred timescale is situated at the first local maximum, which is generally considered where SF slope breaks.
        In contrast, the comparison becomes marginally consistent for sources like NGC 4374 and NGC 4278, and inconsistent for NGC 5077 and NGC 4579.
        We hypothesize that the SF result is strongly biased by the sampling window effect by simulating the actual sampling DRW light curve (cyan dotted light).
        Therefore, we conclude that the SF is not sufficient to constrain the timescale for most of the sources.
        
    \subsection{The Damped Random Walk Model} \label{subsec:drw}
        To quantify the variable behavior of these sources, we have modeled the light curves as the result of a stochastic damped random walk (DRW) process.
        DRW is a specific Markovian stationary Gaussian process. 
        The application of DRW as a mathematical model to describe the optical variability of quasars was first proposed by \citet{2009ApJ...698..895K}.
        Validation of DRW to describe the sub-mm light curve of Sgr A* was then investigated by \citet{2014MNRAS.442.2797D} and applied to other sub-mm sources by \citet{2015ApJ...811L...6B}. 
    	DRW contains only three parameters: the mean of the light curve $\mu$, the long-term stander deviation $\sigma$, and the characteristic timescale $\tau$.
    	The structure function of the DRW model is
    	\begin{equation}
    	    SF^2(\Delta t) = 2 \sigma^2 (1 - e^{-\vert \Delta t/\tau \vert}).
    	\label{eq:sf_drw}
    	\end{equation}
    	The $\tau$ determines where the light curve is decorrelated, which corresponds to the saturation of the variability.
    	In the frequency domain, the PSD of the DRW model can be given as
    	\begin{equation}
    	    \text{PSD}(f) = \frac{4\sigma^2\tau}{1+(2\pi f \tau)^2},
    	\end{equation}
        which is composed of red noise spectrum ($\alpha_{\text{PSD}} = -2$) in the higher frequency regime ($f_\tau \gg 1$) and flat white noise spectrum at low frequencies ($f_\tau \ll 1$).
        	
        We have followed \citet{2009ApJ...698..895K} and \citet{2014MNRAS.442.2797D} to fit light curves with the following procedure.
        The likelihood function of a DRW fitting the observed light curve is given by a product of Gaussian describing the residuals from each light curve point, as
        \begin{equation}
            P(\{x_i\}|\mu, \tau, \sigma) = \prod^{n}_{i=1} [2\pi (\Omega_i+\sigma_{i}^{2})]^{-\frac{1}{2}} \text{exp}[-\frac{1}{2} \frac{(\hat{x}_i-x_i^*)^2}{\Omega_i + \sigma_i^2}], 
        \end{equation}
        where $\sigma_i$ represents measurement uncertainties, and $x_i^*$ is the difference between each data point and mean
        \begin{equation}
            x_i^* = x_i-\mu.
        \end{equation}
        The quantities $\Omega_i$ and $\hat{x}_i$ are iterated by following equations
        \begin{equation}
            \hat{x}_i = a_i\hat{x}_{i-1}+\frac{a_i\Omega_{i-1}}{\Omega_{i-1}+\sigma^2_{i-1}}(x^*_{i-1}-\hat{x}_{i-1})
        \label{eq:6}
        \end{equation}
        and
        \begin{equation}
            \Omega_i = \Omega_1(1-a_i^2) + a^2_i\Omega_{i-1}( 1-\frac{\Omega_{i-1}}{\Omega_{i-1} + \sigma^2_{i-1}} )
        \label{eq:7}
        \end{equation}
        and regulated by $a_i$, which is the expected correlation between data points 
        \begin{equation}
            a_i = \text{exp}[-(t_i-t_{i-1})/\tau].
        \end{equation}
        The iterative procedure starts from the initial conditions 
        \begin{equation}
            \hat{x}_1 = 0, \;\Omega_1=\sigma^2.
        \end{equation}
        Note that $\sigma^2$ is the variance of the light curve.

        \begin{figure*}[ht!]
            \centering
            \includegraphics[width=\textwidth]{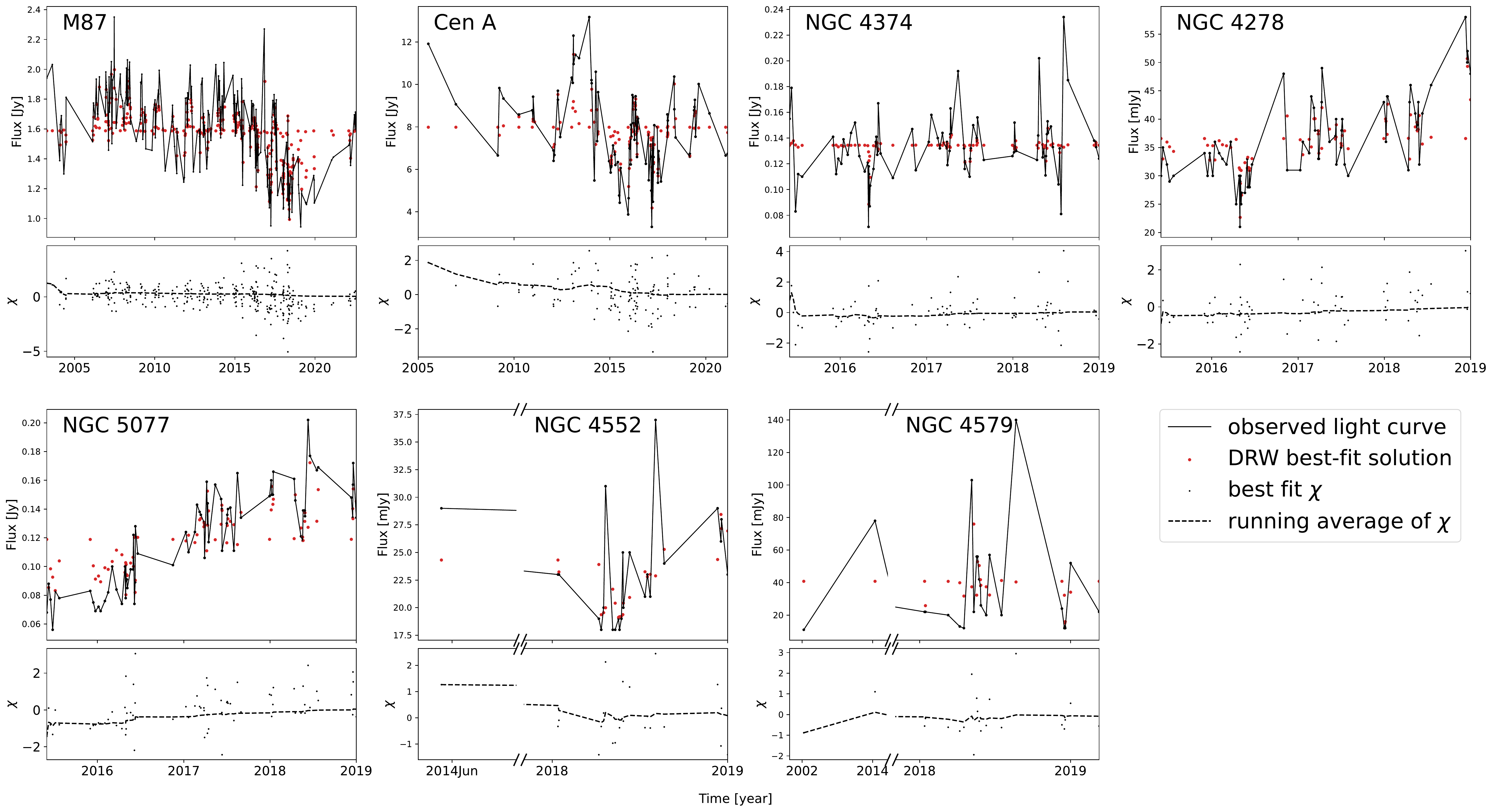}
            \caption{The best-fit DRW solution, the light curve, and the residual $\chi$ along with the time. The running average of the residual is also shown for evaluating the distribution of $\chi$}
            \label{fig:chi}
        \end{figure*}
        
        This procedure allows us to calculate the best-fit DRW solution from a given vector of parameters.
        
        Therefore, the fit quality can be evaluated by examining the distribution of the residual $\chi$ and the reduced chi-square $\chi_n^2$ calculating as
        \begin{equation}
            \chi_{\nu}^2 = \frac{1}{\nu} \sum^n_{i=1} \frac{(\hat{x}_i-x_i^*)^2}{\Omega_i + \sigma^2_i}, 
        \end{equation}
        where $\nu$ is the degree of freedom.
        A proper fit should meet the criteria that the residuals are independent, characterized by a Gaussian distribution with a mean of zero and a variance of one, and consequently the $\chi_{\nu}^2$ should approach to one.
        Figure \ref{fig:chi} illustrates the best-fit DRW solution, light curve, and residual $\chi$ for seven sources.
        The values of $\chi_{\nu}^2$ are listed in the final column of table \ref{table:drw_para}.
        One can observe that the best-fit DRW solution will approach the mean of the light curve if the time interval between two data point is significantly greater than the $\tau$, which agree with the fitting procedure equation \ref{eq:6} and \ref{eq:7}.
        We draw attention to this characteristic of the fitting outcomes, particularly for NGC 4374, since the DRW fits a smaller timescale than the sampling cadence. This results in a DRW light curve with less variability than the actual light curve observed.
        
\begin{figure*}[ht!]
            \centering
            \includegraphics[width=0.6\textwidth]{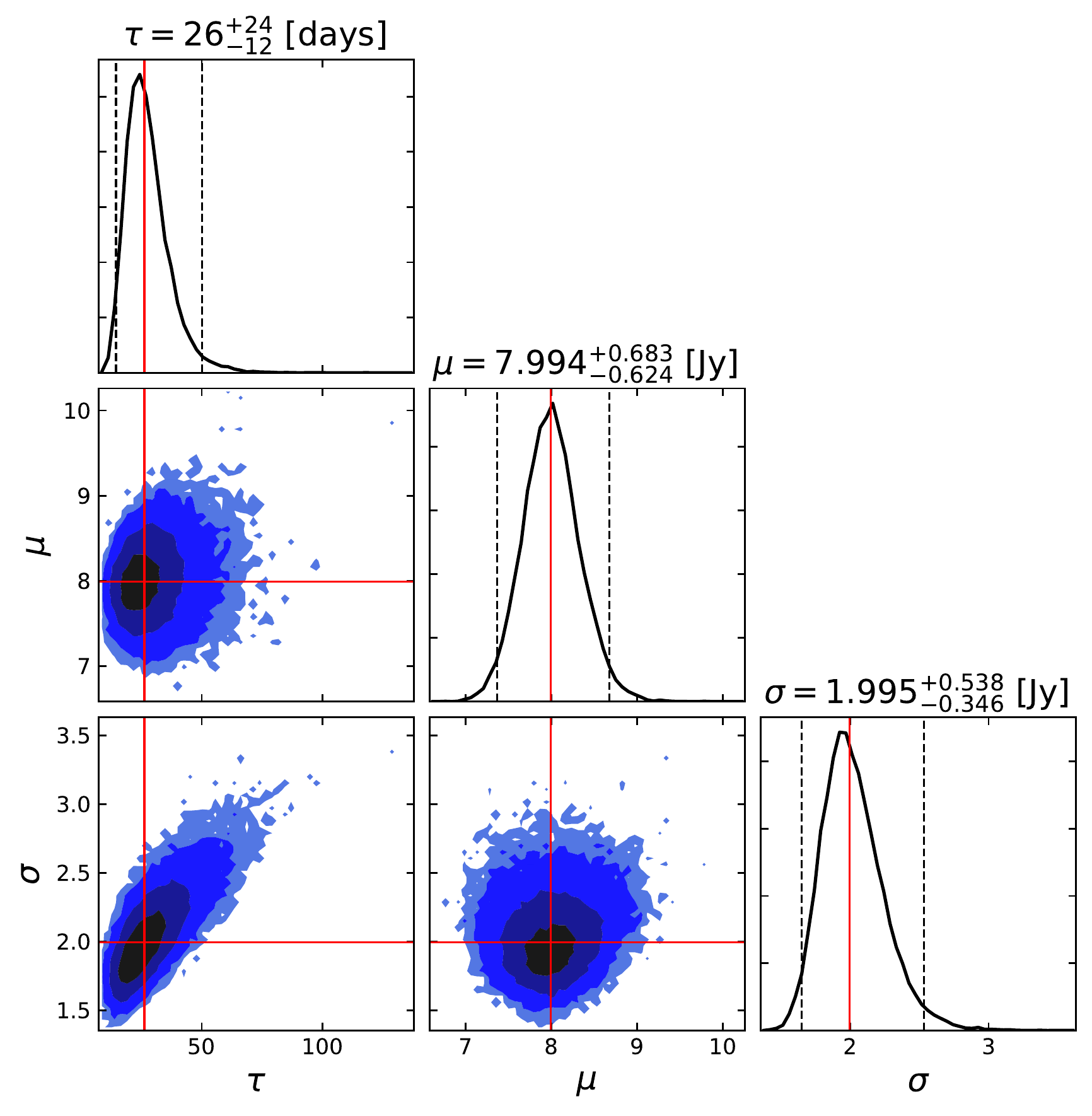}
            \caption{Posterior probability distributions of the DRW model parameters use Cen A as an example. The red vertical line displays the median of samples, and the black dashed line labels the $95 \%$ CI. Contours correspond to 0.2, 0.5, 0.8, and 0.95 of the posterior volume.}
            \label{fig:mcmc}
        \end{figure*}

        We then applied a Bayesian approach to compute the posterior probability of $\tau$, $\mu$, and $\sigma$ from the likelihood.
        Following \citet{2014MNRAS.442.2797D}, we assume uniform priors on all three parameters.
        The Metropolis-Hastings algorithm was adopted to sample $\tau$, $\mu$, and $\sigma$.
        The corner plot of the parameters solution is presented in Figure \ref{fig:mcmc} using Cen A as an example.
        The inferred parameters were estimated by taking the median of the sample population (the red line in figure \ref{fig:mcmc}).
        We adopted a two $\sigma$ uncertainty which corresponds to the 95 $\%$ confidence interval in the population shown in the vertical black dash line.
        
    \subsection{Result}
\begin{figure*}[ht!]
            \centering
            \includegraphics[width=\textwidth]{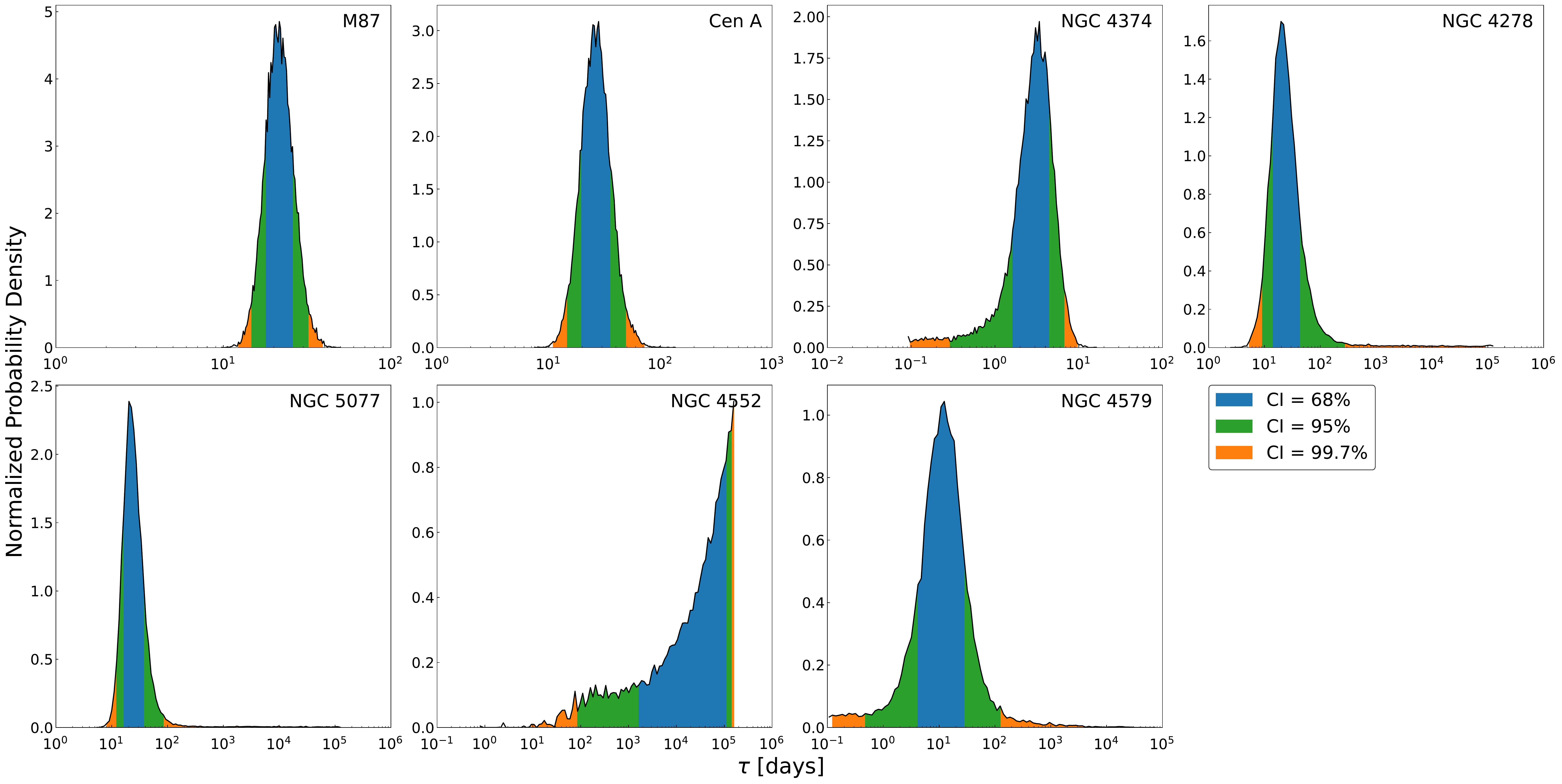}
            \caption{Probability distribution of $\tau$ for each source. The 68, 95, and 99.7$\%$ CI are labeled in blue, green, and orange, respectively.}
            \label{fig:drw}
        \end{figure*}

        We have applied the DRW model to measure the $\tau$ of our six new samples and reanalyze M87.
        Figure \ref{fig:drw} showcases the DRW fit solution of the probability distribution of $\tau$, which is the most relevant parameter in the light curve.
        A thorough examination of the probability distribution was carried out by plotting the 68, 95, and 99.7 $\%$ confidence intervals (CI) for the sample.
        Note that a good measurement is only valid when both sides of the distribution converge at $\tau$ range from $\tau > \Delta t$ and $\tau < T$ where $\Delta t$ is the minimum observation separation in the light curve, and $T$ is the total light curve duration.
        We observe a long tail extending to a short or long timescale, suggesting that $\tau$ of the source is not well captured.
        However, in these cases, the timescale has a high likelihood of being located where the Metropolis-Hastings algorithm conducted intense sampling. 
        As a result, we present the median and 95 $\%$ CI of the distribution to reflect the highest possible parameter value in these sources.
        
    	The inferred timescale $\tau$, the inferred light curve mean $\mu$, and the inferred standard distribution $\sigma$ are presented in Table \ref{table:drw_para}, the reduced-$\chi^2$ of the best-fit result are calculated in the last column.
        We note that the correction for the cosmological time dilation is not necessary in these cases.
        The correction according to Eq.17 in \citet{2009ApJ...698..895K} is $\sim 1 \%$ for the most distant object NGC 5077 ($z \sim 0.00936$; from NED), which is smaller than the inferred uncertainty of the timescale.
    	
    	\begin{table*}[!ht] 
        \centering
        \caption{Inferred parameters of DRW solution. We estimate $\tau$, $\mu$, and $\sigma$ by calculating the median of the parameter probability distribution and estimating the uncertainty using 95 $\%$ CI.}
        \begin{tabular}{c c c c c c c c}
        \hline
        Source & $\tau$ & $\Delta_{\tau}$ & $\mu$ & $\Delta_{\mu}$ & $\sigma$ & $\Delta_{\sigma}$ & $\chi^2$ \\
         & [days] & [days] & [Jy] & [Jy] & [Jy] & [Jy] & \\
        \hline
        M87      & 21   & (15, 32)  & 1.589 & (1.530, 1.648) & 0.2676 & (0.2362, 0.3097) & 1.20\\
        Cen A    & 26   & (14, 50)  & 7.994 & (7.370, 8.677) & 1.9954 & (1.6492, 2.5323) & 1.06\\
        NGC 4278 & 22   & (8, 283)  & 0.037 & (0.032, 0.043) & 0.0077 & (0.0059, 0.0209) & 1.06\\
        NGC 4374 & 3.0  & (0.3, 7)  & 0.134 & (0.128, 0.141) & 0.0250 & (0.0212, 0.0303) & 1.04\\
        NGC 4552 & -    & (70, -)   & -     & (-, -)         & -      & (-, -)            & 1.12\\
        NGC 4579 & 11.4 & (0.4, 49)& 0.040 & (0.011, 0.073) & 0.0386 & (0.0269, 0.1001) & 1.15\\
        NGC 5077 & 24   & (12, 92)  & 0.119 & (0.099, 0.143) & 0.0384 & (0.0300, 0.0683) & 1.07\\
        \hline
        \end{tabular}
        \label{table:drw_para}
        \end{table*}
        
        A long enough monitoring observation is critical for a DRW study, which provides sufficient probe to the long-term variance, i.e., the white noise part of the PDS.
        Combining archive data with our monitoring observation, the total light curve duration for each source is at least ten times longer than the $\tau$ we found, suggesting that the DRW result should be secure in our cases (\citet{2017AA...597A.128K}).
        On the other hand, the stationary character of the underlying physical process becomes a concern for such a long monitoring observation. 
        In other words, the effect of the underlying process change should be considered.
    	
        We found short timescales to be constrained (or with a robust trend to be constrained) in all of our cases except NGC 4552.
        With a proper sampling cadence, both a cadence much shorter than the timescale and a duration much longer than the timescale, we observed that the degree of how well a timescale can be constrained relies on certain amount of data points.
    	For example, in \citet{2014MNRAS.442.2797D}, \citet{2015ApJ...811L...6B} and this study, sources such as Sgr A*, M87, M81, and Cen A, whose light curve has over a hundred data points been accumulated, seems to provide reliable constraint on timescale.
    	NGC 4374, NGC 4278, and NGC 5077, with only around 80 data points be accumulated, exhibit significant uncertainty in either upper or lower limit.
        Furthermore, timescale $\tau$ becomes poorly constrained in the case of NGC 4579 and NGC4552.
    	Therefore, we suggest that a hundred data points seems to be a lower limit number for measureing a reliable timescale in this experiment.
    	Below we discuss the analysis detail for individual cases.
    	
    	\subsubsection{Cen A}
    	We found that the light curve of Cen A can be well described by DRW.
    	A well-constrained $\tau$ was measured, while the assessment from both the distribution of residual and reduced-$\chi^2$ suggest a good fit has been approached.
        However, we measured that there was roughly a two-time difference in $\tau$ when we compared the results of fitting a light curve using linear and logarithmic flux density methods.
    	Except for Cen A, we identify consistent $\tau$ in other sources, similar to the result in \citet{2014MNRAS.442.2797D}.
        Unlike other data, we suspect that Cen A's flux distribution may show a significant trend to be either a Gaussian or a lognormal distribution (\citet{2005MNRAS.359..345U}), and suggests which fit is prefer.
    	We calculated the histogram of the normalized flux of Cen A; however, with limited data, we cannot tell whether Cen A flux follows a Gaussian or Lognormal distribution.
    	Here we adopted the linear result of $\tau=26^{+24}_{-12}$ days since linear flux mean and standard deviation are more straightforward.
    	
    	\subsubsection{NGC 4374}
    	We measured a $\tau=3^{+4}_{-2.7}$ days for NGC4374 with a good fit quality. 
    	A long tail distributed to a short timescale is seen in Figure \ref{fig:drw}, suggesting that the lower limit of NGC4374 is not well constrained.
    	The lower limit long tail reflects the lower limit of 0.3 days for 95 $\%$ CI in Table \ref{table:drw_para}. 
        Note that our highest cadence of observation is about one day.
    	This result suggests that more daily monitoring observations are necessary for capturing NGC4374's high-frequency behavior in order to constrain the timescale.
    	The actual timescale is expected to be larger than three days since the Metropolis-Hastings algorithm will preferentially sample the long tail of the probability distribution rather than the peak.
    	
    	\subsubsection{NGC 4278 $\&$ NGC 5077} 
    	We measured $\tau=22^{+261}_{-14}$ days and $\tau=24^{+68}_{-12}$ days for NGC 4278 and NGC 5077, respectively.
    	In both cases, we observed a long tail at the long timescale, suggesting that we do not have a reasonable upper limit constraint for these sources.
    	They both have $\chi^2 \approx 1$, suggesting a good fit quality. 
        However, the distribution of residual that have a Gaussian distribution deviating from having a zero mean, suggesting that there is additional noise present in the light curve along with the actual DRW signal.
    	In addition, both sources' light curve show a long-term increase trend and lack of a stable flux level.
    	These evidence suggest that our 4-years observation did not capture the long-term variable behavior of the sources, while we still detected some short characteristic timescale of the variability.
    	Furthermore, the fit quality suggests that DRW does not describe the light curve well. 
        A model whose PDS slope index is not equal to -2 (e.g., intraday power spectrum of Sgr A* in \citet{2014MNRAS.442.2797D} and Matérn covariance Gaussian process considered by \citet{2022ApJ...930L..19W}) or is a superposition of multi-stochastic Gaussian is possible to be consider.
    	Based on current results, weekly or monthly monitoring observations every few years may precisely capture the variability properties.
    	
    	\subsubsection{NGC 4579 $\&$ NGC 4552}
    	We have only a small number of data points to marginally constrain the timescale of NGC 4579. 
        Here we adopted a timescale of $\tau=11^{+116}_{-10.6}$ days with $\chi^2=1.15$.
    	The amount of residual is statistically insufficient to provide a robust assessment.
    	In addition, the upper and lower limits are not well constrained, leading to large uncertainty in the inferred  $\tau$.
    	For NGC 4552, we can not constrain any timescale with the current data.
    	Improvement in the numbers of data and flux measurement uncertainty can make a better constraint on timescale measurement in both cases.
    	
    	\subsubsection{M87}
    	We measured a timescale of $\tau = 21^{+11}_{-6}$ days for M87.
    	This measurement is roughly two times smaller than the $\tau$ in \citet{2015ApJ...811L...6B} who measured $\tau = 45^{+61}_{-24}$.
    	To investigate the inconsistency in the $\tau$ measurement, we separated the light curve into two segments using 2015 as the benchmark.
    	The DRW solution for the before 2015 segment, with a $\tau = 36^{+50}_{-20}$ and a $\chi^2 \sim 1$, is consistent with \citet{2015ApJ...811L...6B}; by contrast, the after 2015 segment has a fitting result of $\tau=13^{+10}_{-6}$, $\mu=1.420^{+0.082}_{-0.083}$, and $\sigma=0.105^{+0.25}_{-0.18}$ with a $\chi^2 \sim 1.5$.
        Further, we note that the after 2015 segment is composed of the SMA calibrator database and our observation as the phase calibrator (blue and black marks in Figure \ref{fig:lc}).
    	By fitting them separately, we found that the after 2015 segment's variability properties are dominated by our observation, while we cannot identify any timescale for the SMA calibrator database.
    	We investigate the systematic difference between these two light curves, and we speculate the variability difference comes from the cadence in which our denser weekly to daily observation is advanced to probe the short timescale.
    	
        The discrepancy of the fitting result suggests the variability properties vary over these two segments, possibly indicating an intrinsic physics evolution happens in the historical flux decline, while we have seen a historical low energy state around 2017 to 2018 (also see \citet{2021ApJ...911L..11E}).
        Nevertheless, the flux decline has recovered itself to the mean flux level in recent observation.
    	We point out that whether the underlying process is stationary on such long timescales is essential to consider in the long-term monitoring study.
    	Here, we adopted the result of $\tau \sim 20$, the whole light curve measurement, to represent the averaged M87 variability properties.
    	
    	We noted that M87 shows an extended jet structure in the high SNR sub-mm observation (large-scale jet lobes in an arcsecond scale; see Figure 1 in \citet{2021ApJ...910L..14G}), which we also observed in our M87 snapshot.
    	The core flux can be contaminated by the jet flux in our observation since the angular resolution varies along with the monthly changed configuration schedule.
    	The compact configuration, corresponding to a larger beam size, will naturally include the extended jet component, i.e., the component D in \citet{2021ApJ...910L..14G}.
    	This causes an additional artificial variability to be added to the M87 light curve.
    	To remove this artificial variability, we first investigate the variability of the extended jet structure.
    	From a series of high angular resolution epoch, we found that component D are relevantly stable with a flux of $\sim$ 50 mJy; therefore, we assumed the extended jet structure is static.
    	We note that beam sizes vary from 1 to 10 arcsecond in our observation, and we measured the angular distance between the M87 core and the component D center (here we assumed flux of component D distribute as an elliptical gaussian) to be 3.4 arcsecond from the image in \citet{2021ApJ...910L..14G}.
        We then removed a 50 mJy flux from 66 epochs, whose measurement is considering to include component D base on the beam size and beam position angle.
    	However, the DRW solution of the modified light curve is consistent with the original one, suggesting that DRW is not sensitive to this level of variability in this case.
    	
    	\subsubsection{Limitation $\&$ Future Work}
    	The DRW is useful for quantifying the variability properties of the light curves.
        Moreover, this fitting light curve method shows advantage on getting rid of observation sampling bias comparing to SF method (\citet{2009ApJ...698..895K} and \citet{2014MNRAS.442.2797D}). 
    	However, it naturally limits by itself on the flexibility to probe the light curve properties in the high frequency regime, since the use of DRW assume the PSD slope $\alpha_{\text{PSD}} = -2$.
    	Several studies have indicated a steeper PSD slope in the optical quasar light curve (\citet{2013ApJ...765..106Z} and \citet{2015MNRAS.451.4328K}) and XRB x-ray variability (\citet{2021MNRAS.504.3862T}).
    	In the context of Sgr A* sub-mm light curve, the continuous observing mode using ALMA provide a perfect tool to probe the high frequency regime of the PSD with the use of high cadence high sensitivity light curve, and recently found that $\alpha_{\text{PSD}}$ is steeper than -2 (\citet{2020ApJ...892L..30I}, \citet{2021ApJ...920L...7M} and \citet{2022ApJ...930L..19W}).
    	In \citet{2022ApJ...930L..19W}, a steepening PSD (with measured slope $\alpha_{\text{PSD}}=$ -2 to -3) in short timescale continue with a gentle slope following is found using the SF method.
    	Additionally, they apply the Matérn covariance model, which is also a Gaussian process modeling but provide an extra flexibility on fitting the $\alpha_{\text{PSD}}$, and found that $\alpha_{\text{PSD}}=-2.6$.
    	Compare to the DRW model, the Matérn covariance' $\tau$ measurement is significantly shorter, which is relevant to that DRW can lead bias on $\tau$ measurement when the underlying process has $\alpha_{\text{PSD}}$ not equal to 2 (\citet{2016MNRAS.459.2787K})).
    	A potential caveat of our analysis is that our measured $\tau$ may exist bias under the DRW assumption.
    	The use of the Matérn covariance in Sgr A* was based on high-quality data; hence the application of the Matérn covariance to our irregular sampling light curve, which especially is not sensitive to high frequency, is worth to be explored.
    	On the other hand, to better understand the variability of our targets, specifically the decorrelated timescale and the PSD slope property, both high cadence, high SNR light curve, and longer observation span are necessary.
    	The high-frequency PSD slope has the potential to constrain the theoretical accretion model and expand our understanding of Sgr A* and M87 to a population of low accretion rate systems.

\section{Discussion} \label{sec:discu}
    \subsection{The Black Hole Mass} \label{subsec:BH_mass}
        We have compiled black hole mass from literature in Table \ref{table:bh_mass}.
        Most of the mass were measured from multiple dynamical method, which provided a precise mass estimation.
        By accurately measuring stellar proper motion, Sgr A* has two independent, high precision measurement with a mass of $4.297 \pm 0.012 \times 10^6 M_{\odot}$ from using GRAVITY instrument (\citet{2022AA...657L..12G}) and a mass of $3.951 \pm 0.047 \times 10^6 M_{\odot}$ from using Keck (\citet{2019Sci...365..664D}).
        We adopted a mass of $4.297 \pm 0.012 \times 10^6 M_{\odot}$ for Sgr A* and note that for our purposes the discrepancy in the mass measurement is irrelevant. 
        We adopted a mass of $6.5 \pm 0.7 \times 10^9 M_{\odot}$ for M87, which was determined by fitting the physical parameters of lensed emission near the event horizon (\citet{2019ApJ...875L...6E}).
        This value is consistent with the stellar dynamical method that suggests a mass of $6.6 \pm 0.4 \times 10^9 M_{\odot}$ (\citet{2011ApJ...729..119G}).
        M81 has two mass measurements from both stellar and dynamic method; therefore, we followed \citet{2013ARAA..51..511K} to adopt a mass of $6.5^{+2.5}_{-1.5} \times 10^7 M_{\odot}$ which took the mean of two measurements.
        Cen A has a mass of $5.5 \pm 3 \times 10^7 M_{\odot}$ from the stellar dynamic method (\citet{2009MNRAS.394..660C}).
        For NGC 4374, two groups analyzed the same data but identified different black hole masses.
        We followed \citet{2010ApJ...721..762W} who clarified this discrepancy and adopted a mass of $8.5^{+0.9}_{-0.8} \times 10^8 M_{\odot}$.
        NGC 5077 has a mass of $6.8^{+4.3}_{-2.8} \times 10^8 M_{\odot}$ from the gas dynamic method (\citet{2008AA...479..355D}) and NGC 4552 has a mass of $(4.7 \pm 0.5) \times 10^8$ determined by the stellar dynamic method (\citet{2019ApJ...876..155S}).
	    In addition to dynamical methods, we followed the work done by \citet{2003MNRAS.340..793W} to adopt NGC 4278 and NGC 4579's black hole mass that estimate by $M_{BH}-\sigma$ relationship with a $20 \%$ uncertainty on the stellar velocity dispersion.
        However, we note that systematic errors can be large for all black hole mass estimates.
        
        \begin{table*} \footnotesize
        \caption{Black hole mass compile from literature.}
        \begin{tabular}{c c c c c c}
        \hline
        Source & $M_{BH}$ & Method & Ref & BH shadow size\tablenotemark{a} \\
         & [$M_{\odot}$] &  &  & [$\mu$as]\\
        \hline
        Sgr A*   & $(4.297 \pm 0.012) \times 10^6$   & stellar proper motion & \citet{2022AA...657L..12G}  & $\sim$51.3\\
    	M87      & $(6.5 \pm 0.7) \times 10^9$       & lensed emission       & \citet{2019ApJ...875L...6E} & $\sim$38.2\\
    	M81      & $(6.5^{+2.5}_{-1.5} )\times 10^7$ & stellar $\&$ gas      & \citet{2013ARAA..51..511K}  & $\sim$1.8\\
    	Cen A    & $(5.5 \pm 3) \times 10^7$         & stellar               & \citet{2010PASA...27..449N} & $\sim$1.4\\
        NGC 4278 & $(3.1 \pm 0.5) \times 10^8$       & M-$\sigma$ relation   & \citet{2003MNRAS.340..793W} & $\sim$1.8\\
    	NGC 4374 & $(8.5^{+0.9}_{-0.8}) \times 10^8$ & gas                   & \citet{2010ApJ...721..762W} & $\sim$5.0\\
        NGC 4552 & $(4.7 \pm 0.5) \times 10^8$       & stellar               & \citet{2019ApJ...876..155S} & $\sim$2.8\\
    	NGC 4579 & $(1.1 \pm 0.2) \times 10^8$       & M-$\sigma$ relation   & \citet{2003MNRAS.340..793W} & $\sim$0.6\\
    	NGC 5077 & $(6.8^{+4.3}_{-2.8}) \times 10^8$ & gas                   & \citet{2008AA...479..355D}  & $\sim$1.5\\
        \hline
        \end{tabular}
        \label{table:bh_mass}
        \tablenotetext{a}{We assume a fiducial radius of $5 R_s$ for the potential observable black hole shadow size. The angular diameter of the black hole shadow is calculated by $5 R_s = 10GM/Dc^2$, where $D$ is the distance to the observer. The distances of black hole are adopted from Sgr A*: \citet{2022AA...657L..12G}, M87: \citet{2019ApJ...875L...6E}, M81: \citet{1994ApJ...427..628F}, Cen A:\citet{2010PASA...27..457H}, and others from NASA/IPAC Extragalactic Database (NED).}
        \end{table*}
        
    \subsection{$M_{BH}$ versus $\tau$}
\begin{figure*}[ht!]
            \centering
            \includegraphics[width=0.6\textwidth]{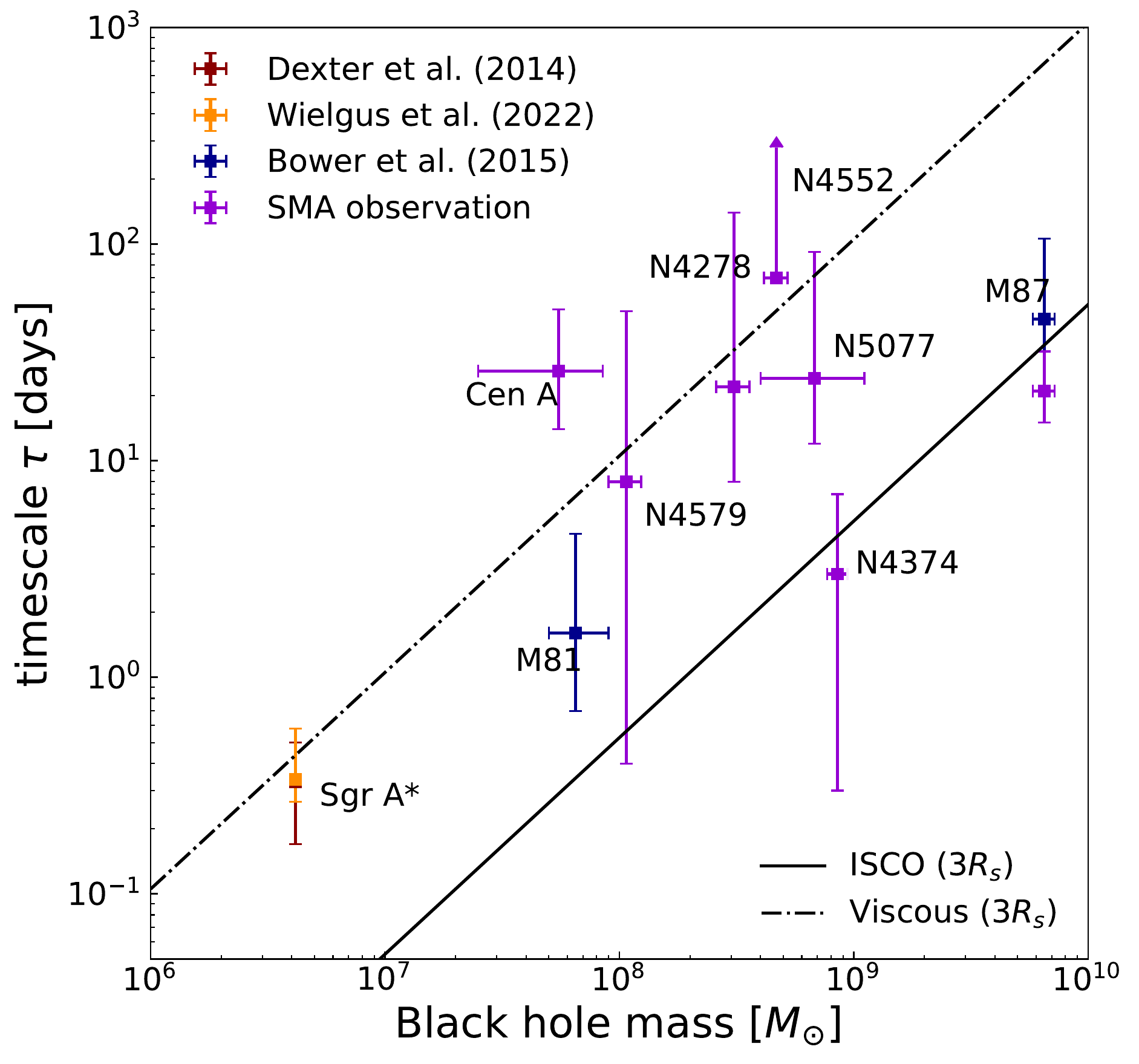}
            \caption{$M_{BH}$-$\tau$ relationship. We present the DRW inferred $\tau$ versus $M_{BH}$ for our six new sources (labeled in purple), as well as the reanalyzed result of M87 (also labeled in purple) and other sources (labeled as different colors according to the provenance) from previous studies.}
            \label{fig:tau_mass}
        \end{figure*}

        To qualitatively understand the physical meaning of the timescale and the connection with the black hole mass, we presented the black hole mass-characteristic timescale correlation among nine sources in Figure \ref{fig:tau_mass}.
        New LLAGN samples in this study are labeled in purple while sources measured in previous studies are labeled in different colors.
        Here we compare these measured timescales with the physical timescales in the accretion system.
        The orbital timescale $t_{orb}$ and the viscous timescale $t_{visc}$, which scale with the black hole mass $M_{BH}$ and Schwarzschild radius $R_s$, can be expressed in following equations:
    	\begin{equation}
    	    t_{orb} \thickapprox 0.5266\bigg{(} \frac{M}{10^8M_{\odot}} \bigg{)} \bigg{(} \frac{R}{3R_s} \bigg{)}^{\frac{3}{2}}\;\text{day},
    	\label{eq:orb}
    	\end{equation}
    	\begin{equation}
    	    t_{visc} \thickapprox 17.55 \bigg{(} \frac{M}{10^8M_{\odot}} \bigg{)} \bigg{(} \frac{R}{3R_s} \bigg{)}^{\frac{3}{2}}\bigg{(} \frac{\alpha}{0.03} \bigg{)}^{-1} \bigg{(} \frac{H/R}{0.6} \bigg{)}^{-2} \; \text{day}.
    	\label{eq:visc}
    	\end{equation}
    	The use of parameter $\alpha \sim 0.03-0.05$ and the scale height $H/R \sim 0.6$ are based on previous simulation result (\citet{2012MNRAS.426.3241N}).
        
    	Two physical timescales are labeled in Figure \ref{fig:tau_mass} with the inner most stable circular orbit (ISCO) radius ($3R_s = 6GM/c^2$) for a Schwarzschild black hole.
    	We found that no source falls under the orbital ISCO timescale, and most sources have a characteristic timescale comparable with the orbital or viscous timescale in the inner region of the accretion disk.
    	Note that the new M87 result agrees with the orbital ISCO timescale, since the period of ISCO varies between 5 days to about a month, relevant to the spin of the black hole from 1 to 0.
        NGC 4374 is situated near the orbital ISCO line. 
        However, as previously discussed in the analysis section, the algorithm has a tendency to choose the unconstrained long tail. 
        As a result, we anticipate that the well-constrained timescale would be larger.

    	\citet{2015ApJ...811L...6B} found a linear $M_{BH}$-$\tau$ correlation among three LLAGNs, Sgr A*, M81, and M87.
    	Our new results agree with the trend they found.
    	Besides the lower limit measurement for NGC 4552, the whole samples show a significant trend that the timescale of these sources increases along with their black hole mass.
    	This can be understood by approximating the timescale scaling around the black hole to be $\tau \varpropto M_{BH}R_s^{3/2}$ (simplification of equations \ref{eq:orb} and \ref{eq:visc}).
    	One can expect a linear black hole mass-minimum timescale correlation to establish when the emission originates from the same $R_s$ where $R_s$ is in the event-horizon scale.
    	In contrast, Cen A shows a much larger timescale in this population and deviates from the linear relationship.
    	The VLBI experiment result (discussed in the next paragraph) suggests that our few-ten-days timescale of Cen A comes from a more outer region of the accretion disk.
        From our result, the black hole mass-minimum timescale  correlation in the sub-mm light curve indicates that sub-mm emission originate from the event-horizon scale for these sources.
    	
    	This interpretation is consistent with our understanding of the VLBI experiment results and the multi-wavelength studies.
    	The synchrotron radiation dominates the emission that originates from the accretion flow at the black hole vicinity, and the synchrotron self-absorption leads to an optically thick photosphere that shrinks along with the frequency until the spectral peak.
    	For instance, we saw the intrinsic size of the emission core shrinks along with the higher frequency VLBI observation for M87 and Sgr A* (\citet{2011Natur.477..185H} and \citet{2018ApJ...865..104J}).
    	The black hole image was then resolved by the Event Horizon Telescope (EHT) in the 2017 campaign using the 230 GHz global VLBI. 
    	The size of the ring is $\sim 5.5 R_s$ and $\sim 4 R_s$, respectively (\citet{2019ApJ...875L...1E} and \citet{2022ApJ...930L..12E}).
    	By contrast, an edge-brightened jet structure was observed in 2017 EHT campaign for Cen A, and the interpreted core size is $\sim 32 \mu$as ($\sim 120 R_s$) (\citet{2021NatAs...5.1017J}).
    	Based on the jet model, optically thin frequency is expected to locate between the wavelengths from Terahertz to far-infrared and further agreed by the Cen A core spectral energy distribution study (\citet{2010ApJ...719.1433A}).
    	In the cases of M81 and NGC 4374, \citet{2018ApJ...853L..14J} and \citet{2021ApJ...922L..16J} report the unresolved core observation with VLBA at 88 GHz.
    	Combined with their flat to inverse slope spectrum, the VLBI measurement suggests a upper limit of core size at 230 GHz to be $< 80 R_s$ and $<30 R_s$ for M81 and NGC 4374.
        On the other hand, the sub-mm emission core size of NGC 4278, NGC 4579, and NGC 5077, lacking of high angular resolution observation, is unknown. 
        However, one can interpret that sub-mm emission is optically thin for NGC 4278 and NGC 4579 from their SED study (see \citet{2013ApJ...777..164M} and \citet{2019MNRAS.490.4606B}).
    	Although the SED study is incompleted, it generally provides properties of the non-thermal emission, and further supports our timescale result.
        
    \subsection{$M_{BH}$-$\tau$ scaling relation}
        To investigate the $M_{BH}$-$\tau$ scaling relationship suggested in Figure \ref{fig:tau_mass}, we perform both linear and power fits on our sample set.
        This sample set includes $\tau$ values from Sgr A* and M81 obtained from literature, and excludes the lower limit constraint of NGC 4552.
        We exclude Cen A from our analysis of the mass scaling relationship because its sub-mm properties, as suggested by VLBI experiments and SED studies, differ from those of our other samples.
        When multiple timescale measurements are available, we use the best measurement for our discussion of the mass scaling relationship. 
        For Sgr A*, we adopt the 8-hour measurement reported in \citet{2022ApJ...930L..19W}, as it provides high-quality data that was previously unattainable.
        Additionally, it is worth mentioning that \citet{2022ApJ...930L..19W} have extensively discussed the comparison of timescale measurements obtained from different sets of light curves, which indicates the evolution of the physical state of Sgr A*.
        For M87, we use our 21-day measurement, as it provides the densest sampling cadence and longest time baseline in our observation.
        
        Following \citet{2007ApJ...665.1489K}, we adopt a Bayesian method and apply a Metropolis-Hastings sampler to account for both $M_{BH}$ and $\tau$ uncertainties in the linear regression process in both linear space and logarithmic space.
        The best fit parameters, i.e., the slope and the intercept, are estimated from the median of the probability distribution, and we present one sigma as the fit uncertainty.

        One can expect a $\tau-M_{BH}$ linear relationship ($\tau \propto M_{BH}R_s^{3/2}$) from the physical interpretation.
        However, we found a linear fit result of $\tau = 2.912^{+0.783}_{-1.030}\; \text{[days]}\; (M_{BH}/10^8M_{\odot})$ with a coefficient of determination of $R^2 \sim 0.23$, suggesting that relationship between $\tau$ and $M_{BH}$ is not simple linear.
        On the other hand, the power-law fit provides a result of
        \begin{equation}
            \tau \approx 3.67^{+1.07}_{-1.56} \; \text{days} \; \bigg( \frac{M_{BH}}{10^8 M_{\odot}}\bigg)^{0.55^{+0.09}_{-0.09}}
    	\label{eq:bestfit}
    	\end{equation}
        with a coefficient of determination of $R^2 \sim 0.61$, and we report a one sigma intrinsic scatter of $0.74^{+0.54}_{-0.3}$ dex for the data.
        
        Based on our limited sample, our experiment suggests that there is a scaling relationship between black hole mass and timescale; however, the correlation is weak and there is significant intrinsic scatter, implying that there may be a third parameter involved in and may not be easy to measured, such as the optical depth.
        We highlight that this relationship is largely driven by Sgr A* and M87. 
        By considering different measurements that reflect different status of the sources' underlying physics for Sgr A* and M87, we found that the relationship can appear more steep or more shallow and tighter or looser.
    	These evidence suggest that the robustness of the relationship requires further samples to be added to the $M_{BH}$-$\tau$ correlation, especially the timescale measurement for black hole mass equal to or smaller than Sgr A* and black hole mass larger than M87.
    	
\begin{figure}[ht!]
            \centering
            \resizebox{\columnwidth}{!}{
            \includegraphics[width=0.6\textwidth]{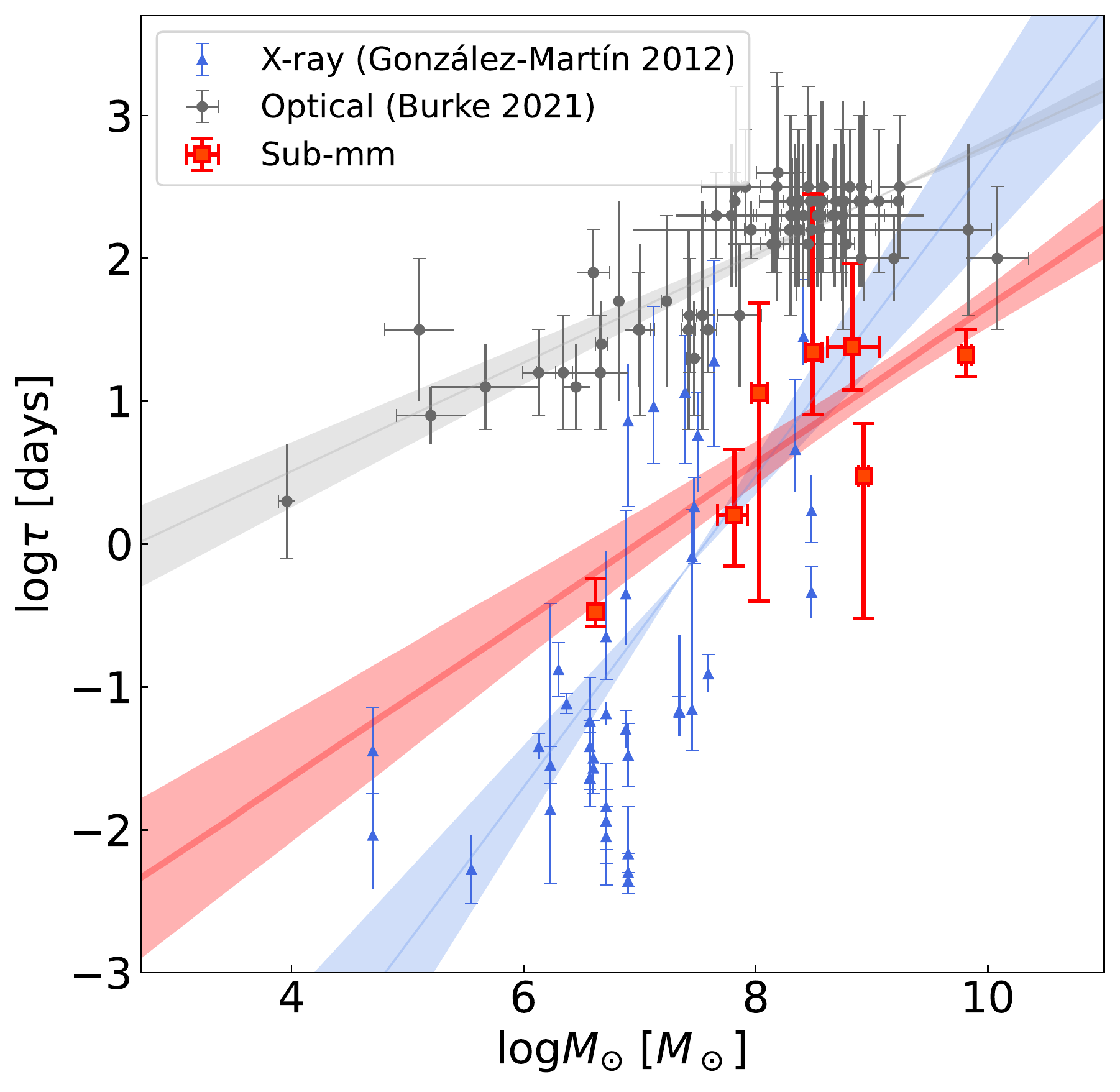}}
            \caption{$M_{BH}$-$\tau$ correlation in sub-mm v.s optical v.s x-ray. Different shapes and colors of marks label data from the different wavelengths. Solid lines represent the linear regression result from each group of data. The shaded area displays the fitting uncertainty in one $\sigma$ level. Note that for sub-mm, the results presented exclude the case of Cen A.}
            \label{fig:tau_mass_fit}
        \end{figure}

    	In Figure \ref{fig:tau_mass_fit}, we present the linear regression result of our sub-mm measurement as well as the optical and the x-ray linear fit from \citet{2021Sci...373..789B} and \citet{2012AA...544A..80G}.
    	Optical data, contains 67 AGNs, has the best fit model of $\tau = 107^{+11}_{-12}\; \text{days}\; (M_{BH}/10^8M_{\odot})^{0.38^{+0.05}_{-0.04}}$ which consider error in both parameters.
    	X-ray data, including 42 light curves measurements, suggest a linear correlation fit of $\tau = 0.02^{+0.02}_{-0.01}\; \text{days}\; (M_{BH}/10^6M_{\odot})^{1.09\pm0.21}$, but without considering the black hole mass error.

    	While various types of AGNs may be classified into different accretion models, such as the standard thin disk model and the radiatively inefficient accretion flow (RIAF), resulting in different wavelengths being attributed to various emission mechanisms, a comparison of variability properties between wavelengths still offers a unique opportunity to gain insight into the disk configuration and the origin of variability.
    	The sub-mm and x-ray wavelengths show a significantly shorter timescale compared to the optical wavelength, suggesting that sub-mm and x-ray variability can provide information about the inner regions of the black hole system. 
        On the other hand, the optical timescale is comparable to that of the outer disk's dynamics or thermal timescale. 
        These observations of the timescales at different wavelengths are in agreement with our current understanding of the structure of the accretion disk/flow.
    	Moreover, slopes are different in the x-ray, sub-mm, and optical $M_{BH}$-$\tau$ correlation, which come with 1, 0.6, and 0.4, respectively.
        In the inner part of the accretion flow, x-ray correlation indicates that plasma around the black hole (corona) correlates tightly with black hole mass.
    	In contrast to x-ray, our sub-mm correlation may be impacted by additional physical parameters such as the optical depth or black hole spin.
        The optical correlation in the outer region suggests that other factors have a significant impact on the variability or that the variability origin differs from that seen in the x-ray or sub-mm data.
        However, it is important to note that the x-ray sample shows more scatter than the optical and sub-mm samples, which was attributed to the absorption along the line of sight proposed by \citet{2018ApJ...858....2G}, indicating a more complex situation.
        Meanwhile, the sub-mm result are less scatter but remains uncertain due to being based on a smaller population.\\

\section{Conclusion} \label{sec:conclu}
    
    In this study, we investigate the variability of seven LLAGNs sub-mm light curves.
    All sources are variable at levels ranging from $\sim 10$ to $\sim 100$\%.
    We show that sub-mm light curves can be well described by DRW model and we determine the characteristic timescales of the variability.
    The LLAGNs' timescales are then found to have a correlation with their black hole mass.
    This correlation suggests that the accretion flow at sub-mm wavelength is transparent down to event horizon scale, and indicate that accretion disk size may scale by black hole mass in low accretion system.
    This method, by measuring timescales, provide an independent way to test the black hole size.
    Moreover, this variability study provides an important identification of potential Event Horizon Telescope targets.
    These sources with their black hole mass and distance to Earth suggest a few microarsecond black hole shadow size, which may be resolvable by the Event Horizon Telescope at a higher frequency (345 or 690 GHz; see \citet{2019BAAS...51g.256D}) or with the long-baselines of space VLBI (e.g., \citet{2020AdSpR..65..821F}, \citet{2021A&A...650A..56R}, and \citet{2022AcAau.196..314G}).

\begin{acknowledgments}
\section*{Acknowledgments}
The Submillimeter Array is a joint project between the Smithsonian Astrophysical Observatory and the Academia Sinica Institute of Astronomy and Astrophysics and is funded by the Smithsonian Institution and the Academia Sinica.
We recognize that Maunakea is a culturally important site for the indigenous Hawaiian people; we are privileged to study the cosmos from its summit.
This paper makes use of the following ALMA data: ADS/JAO.ALMA$\#$2011.0.00010.S, ADS/JAO.ALMA$\#$2012.1.00019.S, ADS/JAO.ALMA$\#$2012.1.00139.S, ADS/JAO.ALMA$\#$2013.1.00803.S, ADS/JAO.ALMA$\#$2013.1.01342.S, ADS/JAO.ALMA$\#$2015.1.00483.S, ADS/JAO.ALMA$\#$2015.1.01577.S, ADS/JAO.ALMA$\#$2017.1.00090.S, ADS/JAO.ALMA$\#$2017.1.00886.L, ADS/JAO.ALMA$\#$2018.A.00062.S. ALMA is a partnership of ESO (representing its member states), NSF (USA) and NINS (Japan), together with NRC (Canada), MOST and ASIAA (Taiwan), and KASI (Republic of Korea), in cooperation with the Republic of Chile. The Joint ALMA Observatory is operated by ESO, AUI/NRAO and NAOJ.
\end{acknowledgments}

\vspace{5mm}
\facilities{SMA, ALMA, PdBI}
\software{MIR-IDL, Miriad, CASA}




\bibliography{sample631}{}
\bibliographystyle{aasjournal}



\end{document}